\documentclass[11pt,a4paper]{article}
\usepackage[hyperref]{emnlp2018}
\usepackage{times}
\usepackage{latexsym}

\usepackage{latexsym}
\usepackage{graphicx}

\usepackage{multicol, latexsym, amsmath, amssymb}
\usepackage{subcaption}
\usepackage{caption}
\usepackage{paralist}

\usepackage[compact]{titlesec}
\aclfinalcopy 

\title{Training and Prediction Data Discrepancies: Challenges of Text Classification with Noisy, Historical Data}

\author{Emilia Apostolova \\
  Language.ai\\
  Chicago, IL USA \\
  {\tt emilia@language.ai} \\\And
  R. Andrew Kreek\\
  Allstate Insurance Company\\
  Seattle, WA USA \\
  {\tt rkree@allstate.com} \\}

\date{}

\begin{document}
\maketitle
\begin{abstract}
  Industry datasets used for text classification are rarely created for that purpose. In most cases, the data and target predictions are a by-product of accumulated historical data, typically fraught with noise, present in both the text-based document, as well as in the targeted labels. In this work, we address the question of how well performance metrics computed on noisy, historical data reflect the performance on the intended future machine learning model input. The results demonstrate the utility of dirty training datasets used to build prediction models for cleaner (and different) prediction inputs.


\end{abstract}

\section{Introduction}

In many benchmark text classification datasets gold-standard labels are available without additional annotation effort (e.g. a star rating for a product review, editorial keywords associated with a news article, etc.). Large, manually annotated datasets for text classification are less common, especially in industry settings. The cost and complexity of a large-scale industry labeling project (with specialized and/or confidential text) could be prohibitive. 

As a result, industry data used for supervised Machine Learning (ML) was rarely created for that purpose. Instead, labels are often derived from secondary sources. For example, text labels may be derived from associated medical billing or diagnosis codes, from an outcome of a litigation, from a monetary value associated with a case, etc. In most cases, the data and labels are a by-product of accumulated historical data. As such, noise is intrinsically present in the data for a variety of incidental reasons, interacting over a long period of time.

In the case of text-based data for document classification, noise could be present in both the text-based document, as well as in the targeted labels. 

There are numerous reasons that could explain the presence of text document noise. For example, industry data based on scanned documents accumulated over time is a common challenge. In some cases, the original image could be lost or unavailable and one is left with the result of OCR engines with varying quality, that could also have changed over time. As various IT personnel handle the data over the years, unaware of its future use, data can be truncated or purged per storage/retention policies. Similarly, character-encoding data transformation bugs and inconsistencies are a common occurrence. In addition, the text data that contains the information needed for correct labelling could be interspersed with irrelevant text snippets, such as system generated messages or human entered notes used for different purposes. 

The reasons for the noise in the targeted labels are also abundant. In cases where the labels are created via human data entry / coding, the reasons could be as mundane as human error or inexperience. In large organizations, department personnel training and management could differ and varying workflows can result in inconsistent labeling. Labeling practices could also evolve over time both at the organization, department, or individual employee levels. Labeling could also be affected by external business reasons. For example, the coding scheme for medical billing codes could have evolved from ICD-9 coding to ICD-10 coding. The billing coding rules themselves could have changed for a variety of accounting and financial reasons, unrelated to the content of the corresponding textual data. Updates in data entry applications could result in a different set of dropdown or checkbox options, and, as a result, coding/labeling could change because of coincidental software updates.

In industry settings, the job of the data scientist often involves exploring, understanding, and utilizing historical datasets for the purposes building prediction models to consume current, and presumably cleaner, document inputs. In such settings, the disparity between the training data and the data used for predictions poses a challenge. Evaluations of various ML approaches should involve performance metrics using not necessarily the available historical training data, but different document inputs. In practice, however, data scientists often ignore the fact that the training data differs substantially from the data the ML model will take as input. Algorithm selections, tuning, and performance metrics are often computed on historical data only. 

In this work, we attempt to address the question of how well performance metrics computed on noisy, historical data reflect the performance on the intended future ML model input. 

\section{Related Work}

In general, the research problem addressed by this work is typically not a concern for strictly academic research relying on benchmark document classification datasets. As a result, relatively few studies address the problem.

Agarwal et al.~\citeyearpar{agarwal2007much} study the effects of different types of noise on text classification performance. They simulate spelling errors and noise introduced through Automatic Speech Recognition (ASR) systems, and observe the performance of Naive Bayes and Support Vector Machines (SVM) classifiers. Agarwal et al. note that, to their surprise, even at 40\% noise, there is little or no drop in accuracy. However, they do not report results on experiments in which the training data is dirty and the test data is clean.

Roy and Subramaniam~\citeyearpar{roy2006automatic} describe the generation of domain models for call centers from noisy transcriptions. They note that successful models can be built with noisy ASR transcriptions with high word error rates (40\%).

Venkata et al.~\citeyearpar{subramaniam2009survey} survey the different types of text noise and techniques to handle noisy text. Similarly to Agarwal et al., they also focus on spelling errors, and on errors due to statistical machine translation. 

Lopresti~\citeyearpar{lopresti2009optical} studies the effects of OCR errors on NLP tasks, such as tokenization, POS tagging, and summarization. Similarly, Taghva et al.~\citeyearpar{taghva2000evaluating} evaluate the effect of OCR errors on text categorization. They show that OCR errors have minimum effect on a Naive Bayes classifier.

All of the above studies focus on text level noise. In contrast, Frenay and Verleyse~\citeyearpar{frenay2014classification} present a survey of classification in the presence of label noise. A number of additional studies focus on techniques improving classifier performance in the presence of label noise by either developing noise-robust algorithms or by introducing pre-processing label cleansing techniques. For example, Hajiabadi et al.~\citeyearpar{hajiabadi2017extending} describe a neural network extended with ensemble loss function for text classification with label noise. Song et al.~\citeyearpar{song2015spectral} describe a refinement technique for noisy or missing text labels. Similarly, Nicholson et al.~\citeyearpar{nicholson2015label} describe and compare label noise correction methods.

More recently, Rolnick et al.~\citeyearpar{rolnick2017deep} investigate the behavior of deep neural networks on image training sets with massively noisy labels, and discover that successful learning is possible even with an essentially arbitrary amount of label noise. 

\section{Method}

To address the question of how well performance metrics computed on dirty, historical data reflect the performance on the intended future ML model input, we evaluated various state-of-the-art document classification algorithms on several document classification datasets, in which noise was gradually and artificially introduced.

\subsection{Types of Noise}
\label{sec-types-noise}

As previously discussed, noise is typically present in historical text-classification training data both within the document texts and within the document labels. To achieve a better understanding of the various types of noise, we evaluated the data present in \textbf{available to us}, historical, industry datasets. With the help of Subject Matter Experts (SMEs), we were able to identify several common types of noise, shown in Table \ref{tab-noise}. In each case, the SME provided a plausible business explanation that justifies the presence of the various types of noise.

\begin{table}[t!]
	\small
\begin{center}
\begin{tabular}{l}

1. Occasionally, documents are replicated and an \\
   identical document could be assigned conflicting labels.  \\
   For example, a single document could describe several \\
   entities, such as multiple participants in a car accident \\
   with document labels derived from their associated \\
   medical billing codes. \\

\hline
2. The difference between a subset of the document  \\
   labels could vary. Some labels could be close to\\
   interchangeable because of various business reasons, \\
   while others are clearly separated. For some labels,\\
   label assignment can be clear-cut and objective, while\\
   for others, human labelers are left to make a subjective \\
   choice.\\

\hline
3. Some documents were truncated as an artifact of the \\
   export process.\\
\hline

4. The information needed to assign correctly a label is \\
   missing from the text document, and instead the human \\
   labeler consulted a different source.\\
\hline

5. There is a large amount of document text irrelevant \\
   to the labeling task at hand, an artifact of the business \\
   workflow and/or export process.\\

\end{tabular}
\end{center}
\caption{\label{tab-noise} Common types of industry noise. }
\end{table}

We then mimicked the described in Table \ref{tab-noise} types of historical data noise in a controlled setting. We gradually introduced noise by randomly flipping a subset of the labels (replicating items 2 and 4 from Table \ref{tab-noise}), replicated a portion of the documents and assigned to them conflicting labels (item 1), truncated the text of some documents (items 3 and 4), and interspersed document with irrelevant text, taken from a different domain (items 4 and 5). All types of noise were introduced gradually and simultaneously, starting from no noise to 100\% noise. 

\subsection{Datasets and Document Classification Algorithms}

\begin{table*}[t!]
	\small
\begin{center}
\begin{tabular}{l|l|l|l|l}
	
 \textbf{Name} &  \textbf{Num. of}  & \textbf{Num. of}  & \textbf{Median Size}  & \textbf{Description}\\ 
   &   \textbf{Documents} & \textbf{Labels} & \textbf{in Tokens} & \\ \hline
20 Newsgroups & 18,828  & 20 & 221 & Messages from Newsgroups on 20 different topics.  \\

\hline
2016 Yelp Reviews & 1,033,124 & 5 & 82 & Yelp reviews dated 2016. The reviews are multi-lingual, \\

& & & & mostly in English. \\

\hline
Synthetic & 115,438 & 5 & 175 & A synthetically created datasets from 5 different document  \\
& & & & collections. Labels correspond to the source collection. \\

\end{tabular}
\end{center}
\caption{\label{tab-0} Summary of the datasets. }
\end{table*}

We focused on several document classification datasets varying in size, in number of training examples, in document length and document content/structure, as well as in the number of label categories. We utilized two common benchmark document classification datasets, and built a third artificial dataset utilizing 5 independent document datasets. Table \ref{tab-0} summarizes the datasets used in our experiments.

The \href{http://qwone.com/~jason/20Newsgroups/}{20 Newsgroups dataset} is a collection of approximately 20,000 newsgroup documents (forum messages), partitioned across 20 different newsgroups. 

The \href{https://www.yelp.com/dataset/challenge}{2016 Yelp reviews} dataset consists of more than 1 million user reviews accompanied with 1 to 5-star business rating (used as document labels). The dataset consists of all available Yelp user reviews dated 2016. 

Both of the above dataset are relatively clean. However, they both rely on user-entered labels. This inevitably leads to some level of noise. For example, in some cases, the content of the user review might not necessarily reflect the user-entered business rating. Similarly, in the case of 20 Newsgroups, a user could send a message to a user group that doesn't necessarily reflect the best message category.

To measure accurately the effects of noise on various algorithm performance, we also created an artificially clean dataset (referred to as \textit{Synthetic}). To create the dataset we utilized 5 different document collections. They include the 20 Newsgroups and a portion of the Yelp reviews dated 2016, described above. We also included the \href{http://www.daviddlewis.com/resources/testcollections/reuters21578/}{Reuters-21578} collection,  a dataset consisting of over 21,000 Reuters news articles from 1987; a Farm Advertisements dataset~\cite{mesterharm2011active} consisting of over 4,000 website text advertisements on various farm animal related topics; a dataset of text abstracts describing National Science Foundation awards for basic research~\cite{Lichman:2013}. The label for each document correspond to the source dataset, i.e. the labels are newsgroup message, review, news article, website ad, v.s. grant abstract. It is trivial for a human annotator to distinguish between the different document categories, and, at the same time, the classification decision involves some understanding of the document text and structure.

In addition, one of our noise-introducing techniques involves interspersing documents with irrelevant text. To introduce irrelevant text snippets in the 20 Newsgroups dataset we utilized texts from the Yelp Reviews dataset and vice versa. To introduce noice in our synthetic dataset we utilized a dataset containing legal cases from the Federal Court of Australia~\cite{galgani2012citation}. 
%
%
%
%
%
%
%

While a thorough comparison of supervised document classification algorithms and architectures is beyond the scope of this work, we experimented with a small number of commonly used in practice document classification algorithms: bag-of-words SVM~\cite{cortes1995support}, a word-level convolutional neural network (CNN) ~\cite{collobert2011natural,kim2014convolutional}, and  fastText~\cite{joulin2016bag}.

\begin{figure*}
  \centering
    \includegraphics[width=0.71\textwidth]{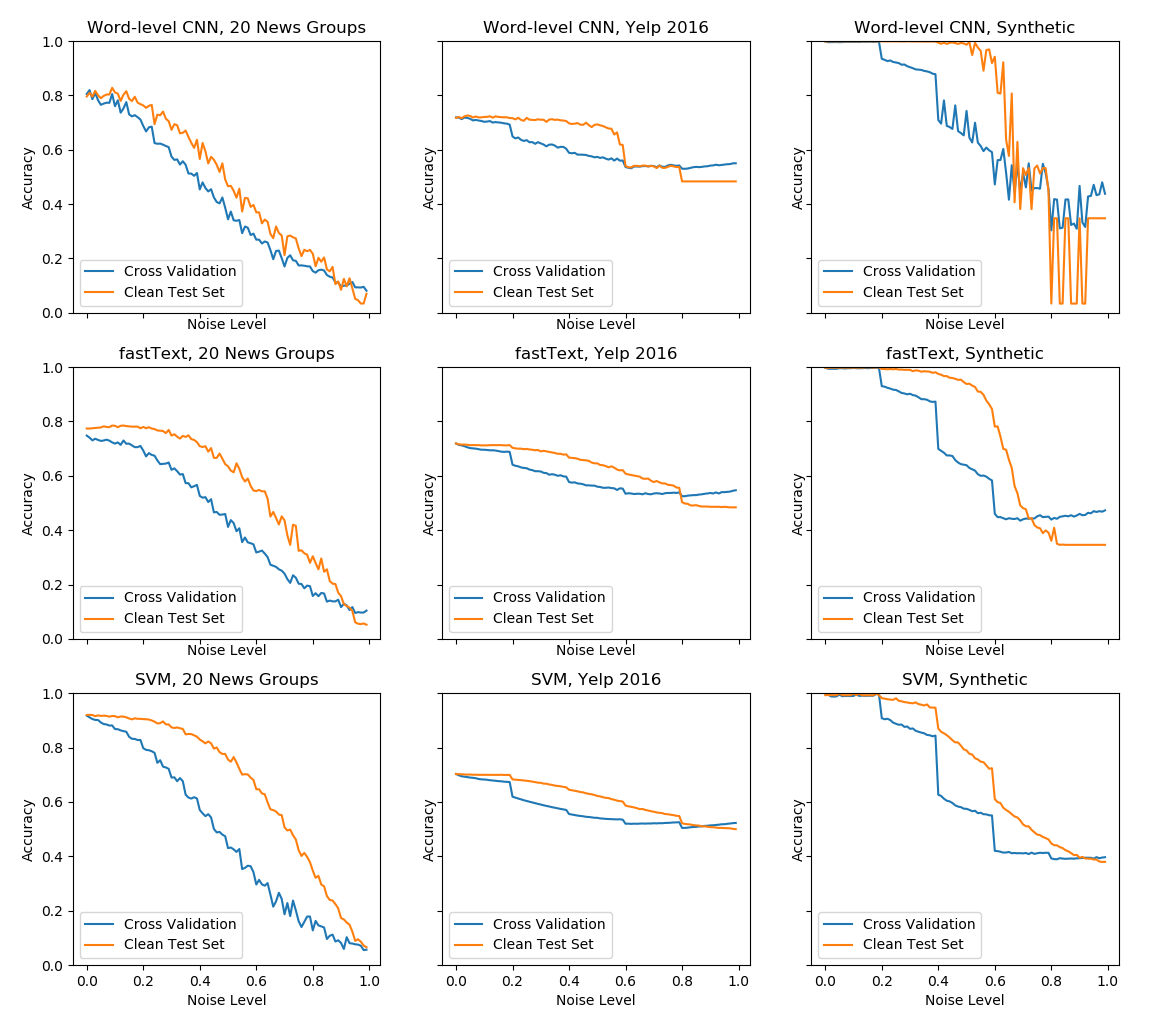}
   \caption{\label{fig1} Word-level CNN, fastText, and bag-of-words SVM performance on the 3 datasets. The y axis shows the model accuracy, as noise is introduced into the training data (x axis). The orange line shows performance on the clean dataset. The blue line shows cross-validation performance measured on the dirty training dataset.}
\end{figure*}

\begin{table*}[t!]
	\small
\begin{center}
\begin{tabular}{l|l|l|l|l|l}
	
\textbf{Algorithm} &  \textbf{Dataset} & \textbf{Slope 0.5} & \textbf{Slope 0.25} & \textbf{Clean set } & \textbf{Clean set }\\
  &   &  &  & \textbf{perf} & \textbf{perf}\\
    &   &  &  & \textbf{gain 0.5} & \textbf{gain 0.25}\\

\hline
CNN & 20Newsgr & -0.62 & -0.27 & 26.15 & 14.71\\
 fastText & 20Newsgr & \textbf{-0.23} & \textbf{0.00} & 35.12 & 14.41\\
SVM & 20Newsgr & -0.28  & -0.09 & 43.09 & 16.30\\
\hline
CNN & Yelp2016 & \textbf{-0.06} & \textbf{-0.04} & 17.25 & 11.44\\
 fastText & Yelp2016 & -0.14 & -0.06 & 13.31 & 10.16\\
SVM & Yelp2016 & -0.17 & -0.08 & 12.95 & 10.97\\
\hline
CNN & Synth & \textbf{-0.01} & \textbf{0.00} & 24.71 & 7.79\\
 fastText & Synthetic & -0.09 & -0.02 & 31.88 & 7.75\\
SVM & Synth & -0.35 & -0.06 & 27.20 & 9.55\\

\end{tabular}
\end{center}
\caption{\label{tab-1} Summary of the clean test dataset performance (orange line in Figure \ref{fig1}). The 3d and 4th columns show the slope of the performance degradation at noise levels 0.5 and 0.25. The fifth and sixth columns show the percentage gain of the performance on the clean test set compared to the dirty training dataset as noise levels 0.5 and 0.25.}
\end{table*}

\subsection{Experiments}
\label{sec-experiments}

We set aside 30\% of the available data from all three datasets as clean test sets. These test sets represents the intended prediction input of the ML model. The rest 70\% of the data was used for training. Noise was gradually introduced into the training sets, which represents dirty, historical training data input that differs from the intended prediction input. 

At each step, the different types of noise (Table \ref{tab-noise}) were introduced, both within the document text and within the document labels: 1) a fraction of the training data texts was truncated by a fraction of the length of the text; 2) a fraction of the training data texts was interspersed with irrelevant text; 3) for a fraction of the categories and a fraction of the texts within each category labels were randomly flipped; 4) for a fraction of the categories, a fraction of the texts were replicated and their labels randomly flipped. For example, at 50\% levels of noise, 50\% of the documents were truncated by 50\% of the length of the text; 50\% of the documents were interspersed with 50\% irrelevant text; for 50\% of the set of the set of labels 50\% of the document labels were randomly flipped; for 50\% of the set of labels, 50\% of the documents were replicated and their labels were randomly flipped. The algorithm performance on the noisy training set was measured via cross-validation. 


In all cases, training was performed without parameter tuning targeting the training dataset (clean or dirty versions). In all cases, text normalization involved only converting the text to lower case. The SVM classifiers were built using unigram bag-of-words, limiting the vocabulary to tokens that appear more than 5 times and in less than 50\% of all documents. For the two smaller dataset (20 Newsgroups and Synthetic) we utilized Wikipedia pre-trained word embeddings of size 100 for both the word-level CNN and fastText classifiers. For the large Yelp Review dataset pre-trained embeddings were not used.  fastText was run using the default training parameters. We experimented with two different word-level CNN architectures both producing comparable results\footnote{Word embeddings of size 100; the sequence length equal the 90th percentile of the training texts; a convolutional layer with 100 filters and window size 8; global max pooling; 0 or 1 dense layers of size 100 and dropout rate of 0.5; ReLU activation; a final dense layer equal to the number of document categories with Softmax activation.}.

\section{Results and Discussion}

Figure \ref{fig1} illustrates the performance of the algorithms on the three datasets. In all cases, the text-classification algorithms demonstrate resilience to noise. The clean dataset performance (orange line) consistently outperforms cross-validation results on the dirty training dataset (blue line)\footnote{We also experimented with different word-level CNN depths on the Yelp 2016 dataset. A deeper architecture (1 additional fully connected dense layer) appears to be slightly more resilient to noise.}.

Figure \ref{fig1} shows performance as noise is introduced into the dataset from 0 to 100\%. In practice, however, dirty historical data used for supervised ML, contains lower levels of noise (otherwise the dataset would be practically unusable). To compare performance of various algorithms in this more realistic setting, we measured the slope of the clean dataset accuracy as noise is introduced from 0 to 0.5, and from 0 to 0.25 (Table \ref{tab-1}). Slope values closer to 0 indicates small performance degradation, while larger negative values correspond to greater performance degradation. Results vary across datasets and algorithms, however, in all cases, the slope of the degradation of performance on a clean test set is small, indicating that all algorithms are able to successfully ignore noise signals at various degrees. The word-level CNN classifier appears to be particularly resilient to relatively small amounts of noise and is the top performer for the 2 larger datasets (Yelp 2016 and Synthetic).

Table \ref{tab-1} also shows the relative performance gain of results on the clean test set compared to results on the dirty training datasets at noise levels 0.5 and 0.25. Results measured on the clean set outperform results measured on the dirty training dataset by an average of 25\% at noise level 0.5 and an average of 11\% at 0.25 noise levels. For the large dataset (Yelp 2016), word-level CNN results show the most significant performance gain.

In addition, results on the artificially clean dataset (Synthetic), demonstrate that practically all the algorithms are almost completely resilient to noise up to 0.5 noise levels. 

\section{Conclusion}

We have shown that text-classification datasets, based on noisy historical data can be successfully used to create high quality prediction models. We analyzed and described types of noise commonly present in historical industry datasets. We simulated simultaneously both text noise and label noise and observed that, across all experiments, the accuracy on a clean dataset significantly outperforms the accuracy measured on the dirty training sets via cross validation. This suggest that traditional accuracy measures on dirty training datasets are typically over-pessimistic. Most remarkably, relatively high noise levels practically have little or no effect on model performance when measured on a clean test set. It could also be extrapolated that artificially created text-classification datasets, e.g. datasets created using a set of imperfect rules or heuristics, could be used to create higher quality prediction models.


\bibliography{noise}
\bibliographystyle{acl_natbib}

\end{document}